\documentclass[aps,pra,onecolumn,groupedaddress,showpacs]{revtex4}

\usepackage{amssymb,amsmath}
\usepackage{graphicx}

\newcommand{\pa}{\partial}

\begin{document}

\title{Two-component breather solution of the nonlinear Klein-Gordon equation}

\author{G. T. Adamashvili}
\affiliation{Technical University of Georgia, Kostava str.77, Tbilisi, 0179, Georgia.\\ email: $guram_{-}adamashvili@ymail.com.$ }

\begin{abstract}
The generalized perturbative reduction method is used to find the two-component vector breather solution of the nonlinear Klein-Gordon equation. It is shown that the nonlinear pulse oscillates with the sum and difference of frequencies and wave numbers in the region of the carrier wave frequency and wave number. Explicit analytical expressions for the profile and parameters of the nonlinear pulse  are obtained. In the particular case, the vector breather coincides with the vector $0\pi$ pulse of self-induced transparency.

\vskip+0.2cm
\emph{Keywords:} Generalized perturbative reduction method,   Nonlinear Klein-Gordon equation, \\ Two-component nonlinear waves.
\end{abstract}

\pacs{05.45.Yv, 02.30.Jr, 52.35.Mw}

\maketitle

\section{Introduction}

A nonlinear coherent interaction of an optical pulse with resonant optical atoms, is governed by the Bloch-Maxwell equations [1,2]. When the Rabi frequency of the wave is real and  the longitudinal and transverse relaxations of the optical atoms are ignored, these equations are reduced to the Sine-Gordon equation [3-6]
\begin{equation}\label{sg}
 \frac{\partial^{2} U}{\partial t^{2}}-C \frac{\partial^{2} U}{\partial z^2} =-\alpha_{0}^{2} \sin U,
\end{equation}
 where $U(z,t)$ is a real function of space coordinate $z$ and time $t$ and represents the wave profile, while  $C$ and $\alpha_{0}^{2}$  are the real constants.

In particular, under the condition of the self-induced transparency, the simplified version of the theory is based on the Sin-Gordon equation, where the function $U$ is the optical pulse envelope [1].

The Sin-Gordon equation also arise in a number of physical areas and applied mathematics. This nonlinear equation was analyzed by means of the inverse scattering transform [4,6], by which it is possible to obtain the complete solution of the equation in the form of the nonlinear solitary waves. Among solitary waves, the nonlinear waves  a relatively low amplitude
\begin{equation}\label{t}
U<<1
\end{equation}
are considered quite often. Under the condition Eq.(2), the Sin-Gordon equation (1) is reduced to the nonlinear (quasilinear) Klein-Gordon equation
\begin{equation}\label{kg}
 \frac{\partial^{2} U}{\partial t^{2}}-C \frac{\partial^{2} U}{\partial z^2} =-\alpha_{0}^{2} U +\frac{\alpha_{0}^{2}}{6} U^{3} -\mathcal{O}( U^5).
\end{equation}

In Ref.[7], was shown that under the conditions of the self-induced transparency, Eq.(3) has the two-component vector $0\pi$ pulse solution. One component oscillates with the sum $w+\Omega_{+}$ ($\kappa+Q_{+}$), and the other with the difference  $w-\Omega_{-}$ ($\kappa-Q_{-})$ of the frequencies (wave numbers) in the region of the parameters $w$ and $\kappa$ that are two or three orders lower than the carrier wave frequency $\omega$ and wave number $k$. In other words, the expressions $\omega/w$ and $k/\kappa$ are of orders $10^{2}\div10^{3}$.  In this case, the conditions of the nonlinear wave existence
\begin{equation}\label{cn1}
\omega>>w>>\Omega_{\pm}>>T^{-1},\;\;\;\;\;\;\;\;\;\;\;\;\;\;\;\;\;\;k>>\kappa>>Q_{\pm}>>T^{-1}
\end{equation}
are fulfilled. Here $\Omega_{\pm}$ and $Q_{\pm}$ are the oscillating parameters, $T$ is the width of the nonlinear pulse. The phase modulation is neglected.

Taking into account the phase modulation  the character of the wave process changes significantly and in Refs.[8-11] were shown that the two-component vector $0\pi$ pulse solution of the Bloch-Maxwell equations  oscillates with the sum and difference frequencies in the region of the carrier wave frequency and wave number of the optical pulse. In this case, the conditions of formation of the two-component vector $0\pi$ pulse have the form
\begin{equation}\label{cn}
\omega>>\Omega_{\pm}>>T^{-1},\;\;\;\;\;\;\;\;\;\;\;\;\;\;\;\;\;\;\;\;\;k>>Q_{\pm}>>T^{-1},
\end{equation}
which are considerably weaker as compared with Eq.(4), considered earlier in Ref.[7].

The purpose of the present work is to consider the two-component vector breather ($0\pi$ pulse) solution of the nonlinear Klein-Gordon equation (3) using the generalized perturbative reduction method [7-18] and obtain two-component vector breather solution in the same conditions Eq.(5) as for the Bloch-Maxwell equation.

The rest of this paper is organized as follows: Section II is devoted to the nonlinear Klein-Gordon equation for slowly varying complex envelope functions. In Section III, using the generalized perturbative reduction method, we will transform Eq.(3) to the coupled nonlinear  Schr\"odinger equations for auxiliary functions.
In Section IV, will be presented the solution of the two-component nonlinear pulse. Finally, in Section V, we will discuss the obtained results.

\vskip+0.5cm

\section{The nonlinear Klein-Gordon equation}

We shall consider situation when a pulse with width $T$, the carrier frequency $\omega$ and wave number $k$, propagating along the positive $z$ axis.
We are interested in the case where the pulse duration is much longer than the inverse frequency of the carrier wave, i.e. $T>>1/\omega$.
Following the standard procedure of the slowly varying envelope approximation, we can transform the nonlinear Klein-Gordon equation (3) for real function $U$ into the slowly varying envelope functions, using the expansion [1,2,19]
\begin{equation}\label{eq2}
U(z,t)=\sum_{l=\pm1}\hat{u}_{l}(z,t) Z_l,
\end{equation}
where $Z_{l}= e^{il(k z -\omega t)}$ is the exponential fast oscillating function, $\hat{u}_{l}$ are the slowly varying complex envelope functions, which satisfied inequalities
\begin{equation}\label{swa}\nonumber
 \left|\frac{\partial \hat{u}_{l}}{\partial t}\right|\ll\omega |\hat{u}_{l}|,\;\;\;\;\;\;\;\;\;\;\;
 \left|\frac{\partial \hat{u}_{l}}{\partial z }\right|\ll k|\hat{u}_{l}|.
\end{equation}
For the reality of $U$, it is supposed that the expression $ \hat{u}_{+1}= \hat{u}^{*}_{-1}$  is valid.

The linear Klein-Gordon equation is given by
\begin{equation}\label{lin}
\frac{\pa^{2} U}{\pa t^2} -C \frac{\pa^{2} U}{\pa z^{2}}=-\alpha_{0}^{2} U.
\end{equation}

Substituting  Eq.(6) into (7) we obtain the dispersion relation
\begin{equation}\label{dis}
{\omega}^{2}= C k^{2}+\alpha_{0}^{2},
\end{equation}
and the rest part of the linear equation (7) in the form
\begin{equation}\label{lin2}\nonumber
\sum_{l=\pm1}Z_l (-2il\omega \frac{\partial \hat{u}_{l}}{\partial
t}- 2ilkC \frac{\partial \hat{u}_{l}}{\partial z}+\frac{\partial^{2} \hat{u}_{l}}{\partial t^{2}} -C
\frac{\partial^{2} \hat{u}_{l}}{\partial z^{2}})=0.
\end{equation}

The nonlinear Klein-Gordon equation contains the nonlinear term
$$
\sum_{l=\pm1}Z_l \hat{u}_{+1}\hat{u}_{-1}\hat{u}_{l}.
$$
As a results, Eq.(3) for the complex envelope functions $\hat{u}_{l}$ has the form
\begin{equation}\label{nl}
\sum_{l=\pm1}Z_l (-2il\omega \frac{\partial \hat{u}_{l}}{\partial
t}- 2ilkC \frac{\partial \hat{u}_{l}}{\partial z}+\frac{\partial^{2} \hat{u}_{l}}{\partial t^{2}} -C
\frac{\partial^{2} \hat{u}_{l}}{\partial z^{2}}-\hat{u}_{+1}\hat{u}_{-1}\hat{u}_{l})=0.
\end{equation}

\vskip+0.5cm

\section{The generalized perturbative reduction method}

In order to consider the two-component vector breather solution of the Eq.(3), we use the generalized perturbative reduction method [7-18] which makes it possible to transform the nonlinear Klein-Gordon  equation for the functions $\hat{u}_{l}$ to the coupled nonlinear Schr\"odinger  equations for auxiliary functions.
As a result, we obtain a two-component nonlinear pulse oscillating with the  difference and sum of the frequencies and wave numbers. 

In the frame of this method, the complex envelope function  $\hat{u}_{l}$ can be represented as
\begin{equation}\label{gprm}
\hat{u}_{l}(z,t)=\sum_{\alpha=1}^{\infty}\sum_{n=-\infty}^{+\infty}\varepsilon^\alpha
Y_{l,n} f_{l,n}^ {(\alpha)}(\tau, \zeta_{l,n}),
\end{equation}
where $\varepsilon$ is a small parameter,
$$
Y_{l,n}=e^{in(Q_{l,n}z-\Omega_{l,n}
t)},\;\;\;\;\;\;\;\;\;\;\;\;\;\;\;\;\;\;\zeta_{l,n}=\varepsilon Q_{l,n}(z-v_{{g;}_{l,n}} t),
$$
$$
\tau=\varepsilon^2 t,\;\;\;\;\;\;\;\;\;\;\;\;\;\;\;
v_{{g;}_{l,n}}=\frac{\partial \Omega_{l,n}}{\partial Q_{l,n}}.
$$

It is assumed that the quantities $\Omega_{l,n}$, $Q_{l,n}$ and $f_{l,n}^{(\alpha)}$ satisfies the inequalities for any $l$ and $n$:
\begin{equation}\label{rtyp}\nonumber\\
\omega\gg \Omega_{l,n},\;\;\;\;\;\;\;\;\;\;\;\;k\gg Q_{l,n},\;\;\;
\end{equation}
$$
\left|\frac{\partial f_{l,n}^{(\alpha )}}{ \partial t}\right|\ll \Omega_{l,n} \left|f_{l,n}^{(\alpha)}\right|,
\;\;\;\;\;\;\;\;\;\;\;\;\;
\left|\frac{\partial f_{l,n}^{(\alpha )}}{\partial z}\right|\ll Q_{l,n} \left|f_{l,n}^{(\alpha )}\right|.
$$

Substituting Eq.(10) into (9), for the nonlinear  Klein-Gordon  equation we obtain
\begin{equation}\label{eqz}
\sum_{l=\pm1}\sum_{\alpha=1}^{\infty}\sum_{n=\pm 1}\varepsilon^\alpha Z_{l} Y_{l,n}[W_{l,n}
+\varepsilon J_{l,n} \frac{\partial }{\partial  \zeta_{l,n}} - \varepsilon^2 i l h_{l,n}  \frac{\partial }{\partial \tau}
-\varepsilon^{2} Q^{2} H_{l,n}\frac{\partial^{2} }{\partial \zeta^{2}_{l,n}}+O(\varepsilon^{3})]f_{l,n}^{(\alpha)}$$$$=
\sum_{l=\pm1}Z_l \hat{u}_{+1}\hat{u}_{-1}\hat{u}_{l},
\end{equation}
where
\begin{equation}\label{cof}
W_{l,n}=- 2 n l\omega \Omega_{l,n}  + 2  n  l k Q_{l,n} C - \Omega^{2}_{l,n}+C  Q_{l,n}^{2},
$$
$$
J_{l,n}=2i Q_{l,n} [l \omega  v_{{g;}_{l,n}}   -l k C   + n \Omega_{l,n}  v_{{g;}_{l,n}}   -C n Q_{l,n}],
$$
$$
h_{l,n}=2(\omega + ln \Omega_{l,n}),
$$
$$
H_{l,n}=   C- v_{{g;}_{l,n}}^{2}.
\end{equation}

Equating to zero, the terms with the same powers of $\varepsilon$, from the Eq.(11) we obtain a series of equations. In the first order of $\varepsilon$, we have
a connection between of the parameters $\Omega_{l,n}$ and $Q_{l,n}$. When
\begin{equation}\label{fo1}
2 ( C k Q_{\pm1, \pm1} -\omega \Omega_{\pm1, \pm1}) - \Omega^{2}_{\pm1, \pm1} + C  Q_{\pm1, \pm1}^{2}=0,
\end{equation}
than $ f_{\pm1, \pm1}^{(1)}\neq0$  and when
\begin{equation}\label{fo2}
2 ( C k Q_{\pm1, \mp1} -\omega \Omega_{\pm1, \mp1}) + \Omega^{2}_{\pm1, \mp1} - C  Q_{\pm1, \mp1}^{2}=0,
\end{equation}
than $ f_{\pm1, \mp1}^{(1)}\neq0$.

From Eq.(12), in  the second order of $\varepsilon$, we obtain the equation
\begin{equation}\label{jo}\nonumber
J_{\pm1, \pm1}=J_{\pm1, \mp1}=0
\end{equation}
and the expression
\begin{equation}\label{v}\nonumber\\
v_{{g;}_{l,n}}=C \frac{  k + l n Q_{l,n} }{ \omega   +l n  \Omega_{l,n} }.
\end{equation}

Substituting Eqs.(6) and (10) into (3), for the nonlinear part of the Klein-Gordon  equation we obtain
\begin{equation}\label{non}
Z_{+1} \frac{\alpha_{0}^{2}}{2} [ ( | f_{+1,+1}^ {(1)}|^{2} +2 | f_{+1,-1}^ {(1)}|^{2} ) Y_{+1,+1} f_{+1,+1}^ {(1)}+  ( | f_{+1,-1}^ {(1)}|^{2}   +2   | f_{+1,+1}^ {(1)}|^{2} )Y_{+1,-1} f_{+1,-1}^ {(1)} ]
\end{equation}
and plus terms proportional to $ Z_{-1}$.

Taking into account the Eqs.(11) and (15), in the third order of $\varepsilon$  the nonlinear Klein-Gordon  equation  is given by
\begin{equation}\label{lk}
\sum_{n=\pm 1} Y_{+1,n}[ - i  h_{+1,n}  \frac{\partial }{\partial \tau}- Q_{+1,n}^{2} H_{+1,n}\frac{\partial^{2} }{\partial \zeta^{2}_{+1,n}}]f_{+1,n}^{(1)}=$$$$
\frac{\alpha_{0}^{2}}{2} [ ( | f_{+1,+1}^ {(1)}|^{2} +2 | f_{+1,-1}^ {(1)}|^{2} ) Y_{+1,+1} f_{+1,+1}^ {(1)}+  ( | f_{+1,-1}^ {(1)}|^{2}   +2   | f_{+1,+1}^ {(1)}|^{2} )Y_{+1,-1} f_{+1,-1}^ {(1)} ].
\end{equation}
Analogous equations proportional to $Z_{-1}$.

From Eq.(16), for the nonlinear Klein-Gordon  equation (3), we obtain the system of nonlinear equations
\begin{equation}\label{2eq}
  i \frac{\partial f_{+1,+1}^{(1)}}{\partial \tau} + Q_{+1,+1}^{2} \frac{H_{+1,+1} }{h_{+1,+1}} \frac{\partial^2 f_{+1,+1}^{(1)}}{\partial \zeta_{+1,+1} ^2}
  +\frac{\alpha_{0}^{2}}{2  h_{+1,+1}} ( | f_{+1,+1}^ {(1)}|^{2} + 2 | f_{+1,-1}^ {(1)}|^{2} ) f_{+1,+1}^ {(1)}=0,
$$$$
i \frac{\partial f_{+1,-1 }^{(1)}}{\partial \tau} + Q_{+1,-1}^{2} \frac{H_{+1,-1} }{h_{+1,-1}} \frac{\partial^2 f_{+1,-1 }^{(1)}}{\partial \zeta_{+1,-1}^2} +\frac{\alpha_{0}^{2}}{2  h_{+1,-1}}( |f_{+1,-1}^ {(1)}|^{2} +2 |f_{+1,+1} ^ {(1)}|^{2} )  f_{+1,-1}^ {(1)}=0.
 \end{equation}

Taking into account Eqs.(6) and (10), after transformation back to the space coordinate $z$ and time $t$, from the system of equations (17) we obtain the coupled nonlinear Schr\"odinger equations for the auxiliary functions $\Lambda_{\pm}=\varepsilon  f_{+1,\pm1}^{(1)}$ in the following form
\begin{equation}\label{pp2}
i (\frac{\partial \Lambda_{\pm}}{\partial t}+v_{\pm} \frac{\partial  \Lambda_{\pm}} {\partial z}) + p_{\pm} \frac{\partial^{2} \Lambda_{\pm} }{\partial z^{2}}
+q_{\pm}|\Lambda_{\pm}|^{2}\Lambda_{\pm} +r_{\pm} |\Lambda_{\mp}|^{2} \Lambda_{\pm}=0,
\end{equation}
where
\begin{equation}\label{OmQ}
p_{\pm}=\frac{ C- v_{\pm}^{2} }{2(\omega \pm \Omega_{\pm})},
$$
$$
 q_{\pm}=\frac{ \alpha_{0}^{2}}{4 (\omega \pm \Omega_{\pm})},
$$
$$
r_{\pm}=2 q_{\pm},
$$
$$
v_{\pm }= v_{g;_{+1,\pm 1}}=C \frac{  k \pm  Q_{\pm} }{ \omega   \pm   \Omega_{\pm } },
$$
$$
\Omega_{+}=\Omega_{+1,+1}= \Omega_{-1,-1},
$$
$$
\Omega_{-}= \Omega_{+1,-1}= \Omega_{-1,+1},
$$
$$
Q_{+}=Q_{+1,+1}= Q_{-1,-1},
$$
$$
Q_{-}= Q_{+1,-1}= Q_{-1,+1}.
\end{equation}

\vskip+0.5cm

\section{The solution of the nonlinear Klein-Gordon  equation}

The solution of Eq.(18) is given by [7-14]
\begin{equation}\label{19}
\Lambda_{\pm }=\frac{A_{\pm }}{T}Sech(\frac{t-\frac{z}{V_{0}}}{T}) e^{i(k_{\pm } z - \omega_{\pm } t )},
\end{equation}
where $A_{\pm },\; k_{\pm }$ and $\omega_{\pm }$ are the real constants, $V_{0}$ is the velocity of the nonlinear wave. We assume that
$k_{\pm }<<Q_{\pm }$  and $\omega_{\pm }<<\Omega_{\pm }.$

Combining Eqs.(6), (10) and (20), we obtain the two-component vector breather solution of the nonlinear Klein-Gordon  equation (3) in the following form:
\begin{equation}\label{vb}
U(z,t)=\mathfrak{A} Sech(\frac{t-\frac{z}{V_{0}}}{T})\{  \cos[(k+Q_{+}+k_{+})z -(\omega +\Omega_{+}+\omega_{+}) t]
$$$$
+(\frac{p_{-}q_{+}-p_{+}r_{-}} {p_{+}q_{-}- p_{-}r_{+}})^{\frac{1}{2}} \cos[(k-Q_{-}+k_{-})z -(\omega -\Omega_{-}+\omega_{-})t]\},
\end{equation}
where $\mathfrak{A}$ is amplitude of the nonlinear pulse. The expressions for  the parameters $k_{\pm }$ and $\omega_{\pm }$  are given by
\begin{equation}\label{rt16}
k_{\pm }=\frac{V_{0}-v_{\pm}}{2p_{\pm}},
$$$$
\omega_{+}=\frac{p_{+}}{p_{-}}\omega_{-}+\frac{V^{2}_{0}(p_{-}^{2}-p_{+}^{2})+v_{-}^{2}p_{+}^{2}-v_{+}^{2}p_{-}^{2}
}{4p_{+}p_{-}^{2}}.
\end{equation}

In the particular case, when function $U$ is  the area of the optical pulse envelope the vector breather solution Eq.(21) coincides with the vector $0\pi$ pulse of self-induced transparency.

\vskip+1.5cm

\section{Conclusion}

We investigate the two-component vector breather solution of the nonlinear Klein-Gordon equation (3). For the nonlinear pulse with the width $T>>\Omega_{\pm }^{-1}>>\omega^{-1}$  used the slowly varying envelope approximation (6) and as results, have been obtained Eq.(9).

Using the generalized perturbative reduction method (10), the nonlinear Klein-Gordon equation is transformed to the coupled
nonlinear Schr\"odinger equations (18)  for the auxiliary functions $\Lambda_{\pm 1}$. It is shown that, under these conditions, one can obtain two-component vector breather (21). The components of this pulse oscillate with the sum $\omega +\Omega_{+}$ $(k+Q_{+})$ and the difference $\omega -\Omega_{-}$ $(k-Q_{-})$ of the frequencies (wave numbers) in the region of the carrier wave frequency $\omega$ and wave number $k$. The dispersion relation and the connection  between parameters $\Omega_{\pm}$ and $Q_{\pm}$ are determined from Eqs.(8), (13) and (14). The parameters of the pulse from Eqs.(19) and (22) are determined.  Under the condition of self-induced transparency the function $U$ is area of the pulse envelope and the vector breather coincides with the vector $0\pi$ pulse of self-induced transparency.

Summarizing the above results, we see that the two-component vector breather  oscillating with the sum and difference of the frequencies and wave numbers (vector $0\pi$ pulse of self-induced transparency) Eq.(21) for the nonlinear Klein-Gordon equation can be formed under the condition of Eq.(5)
in the region of the carrier wave frequency and wave number, which is considerably weaker as compared with the conditions of the existence of the nonlinear wave Eq.(4) considered earlier for the nonlinear Klein-Gordon equation (3) in Ref.[7].

Although we consider the nonlinear Klein-Gordon equation for optics, the obtained results are valid for any other fields of physics where are used this equation.

\vskip+2.5cm

\end{document}